\documentclass[aps,prc,floatfix,showpacs,twocolumn, preprintnumbers]{revtex4-2}

\usepackage{caption}
\usepackage{subcaption}
\usepackage{graphicx}
\usepackage{amssymb}
\usepackage{amsmath}
\usepackage{cancel}
\usepackage{placeins}
\usepackage[dvipsnames,usenames]{xcolor}
\usepackage{dcolumn}
\usepackage{bm}
\usepackage{physics}
\usepackage{hyperref}
\usepackage{float}
\usepackage{soul}

\usepackage[utf8]{inputenc}

\usepackage{bm}
\usepackage{color}
\usepackage{graphicx,epsfig}
\usepackage{amsfonts,amssymb,amsmath}
\usepackage{float}
\usepackage{xcolor}
\usepackage{color}
\usepackage{multirow}
\usepackage{subcaption}
\usepackage{soul}

\usepackage{dcolumn}     
\newcommand{\nuc}[2] {$^{#1}$#2}
\newcommand{\nc}{\newcommand}

\newcommand{\beq}{\begin{equation}}
\newcommand{\eeq}{\end{equation}}
\nc{\bfx}{{\bf x}}
\nc{\bfy}{{\bf y}}
\nc{\bfz}{{\bf z}}
\nc{\bfxh}{{\bf \hat{x}}}
\nc{\bfyh}{{\bf \hat{y}}}
\nc{\bfzh}{{\bf \hat{z}}}
\nc{\bfj}{{\bf j}}
\nc{\bfr}{{\bf r}}
\nc{\bfR}{{\bf R}}
\nc{\bfk}{{\bf k}}
\nc{\bfq}{{\bf q}}
\nc{\bfp}{{\bf p}}
\nc{\bfv}{{\bf v}}
\nc{\bfs}{{\bf s}}
\nc{\bfA}{{\bf A}}
\nc{\bfJ}{{\bf J}}
\nc{\bfsg}{{\bm \sigma}}
\nc{\bfta}{{\bm \tau}}
\nc{\bfvh}{{\bf \hat{v}}}
\nc{\bfqh}{{\bf \hat{q}}}
\nc{\low}{\delta_{\rm Low}}

\nc{\swap}{\rightleftharpoons}

\def\beq{\begin{equation}}
\def\eeq{\end{equation}}
\def\beqy{\begin{eqnarray}}
\def\eeqy{\end{eqnarray}}

\begin{document}

\preprint{LA-UR-24-32074}

\title{Electromagnetic radii of light nuclei from variational Monte Carlo calculations}

\author{G. B. \ King$^{1}$}
\email{kingg@lanl.gov}
\author{G. \ Chambers-Wall$^{2}$}
\email{chambers-wall@wustl.edu}
\author{A. \ Gnech$^{3,4}$}
\email{agnech@odu.edu}
\author{S.\ Pastore$^{2,5}$}
\email{saori@wustl.edu}
\author{M.\ Piarulli$^{2,5}$}
\email{m.piarulli@wustl.edu}
\author{\mbox{R. B.\ Wiringa$^6$}}
\email{wiringa@anl.gov}

\affiliation{
$^1$\mbox{Theoretical Division, Los Alamos National Laboratory, Los Alamos, NM 87545, USA}\\
$^2$\mbox{Department of Physics, Washington University in Saint Louis, Saint Louis, MO 63130, USA}\\
$^3$\mbox{Department  of  Physics,  Old  Dominion  University,  Norfolk,  VA  23529}\\
$^4$\mbox{Theory  Center,  Jefferson  Lab,  Newport  News,  VA  23610}\\
$^5$\mbox{McDonnell Center for the Space Sciences at Washington University in St. Louis, MO 63130, USA}\\
$^6$\mbox{Physics Division, Argonne National Laboratory, Argonne, IL 60439}\\
}

\begin{abstract}
We present variational Monte Carlo calculations of charge and magnetic radii in $A \le 10$ nuclei. The calculations are based on the Norfolk two- and three-nucleon interactions, and associated one- and two-nucleon electromagnetic charge and current operators derived up to next-to-next-to-next-to leading order in the chiral expansion. The charge and magnetic radii are extracted from the respective form factors. We find that the charge radii are within 5\% of the experimental values for the nuclei considered. For the magnetic radii, a comparison is available only with $^3$H and $^3$He electron scattering data that are affected by large error bars. We hope that our predictions foster an interest in precisely measuring magnetic radii of heavier systems.
\end{abstract}

\maketitle

{\it Introduction and Conclusions.} -- Recently, the authors of the present study reported many-body calculations of magnetic moments and electromagnetic form factors in light nuclei based on a chiral effective field theory approach that retains pions, nucleons, and $\Delta$'s as relevant degrees of freedom~\cite{Chambers-Wall:2024fha,Chambers-Wall:2024uhq,king:2024}. Specifically, we adopted the Norfolk  two- and three-nucleon local potentials~\cite{Piarulli:2014bda,Piarulli:2016vel,Piarulli:2017dwd,Baroni:2018fdn,Piarulli:2019cqu}, and their associated many-body electroweak charge and current operators~\cite{Pastore:2008ui,Pastore:2009is,Pastore:2011ip,Piarulli:2012bn,Schiavilla:2018udt}. The calculations covered a broad range of kinematics, from the low-energy structure associated with the nuclear magnetic moments to high-momentum transfer elastic electron scattering. In addition to finding a favorable comparison with the available experimental data, we studied in detail the effects of two-body currents, the contribution of higher-order multipoles, and the sensitivity of  our predictions to variations in the parameters of the interaction models.

In this work, we investigate charge and magnetic radii in light systems, using the Norfolk IIb$^{\star}$~\cite{Piarulli:2014bda,Piarulli:2016vel,Piarulli:2017dwd,Baroni:2018fdn,Piarulli:2019cqu} model as the basis of our analysis. Experimental charge radii are available for both stable and unstable systems, offering additional constraints on the Norfolk interactions and ensuing density distributions, and the study of magnetic radii provides further insight into the low-energy magnetic structure of these systems. 
For the charge radii obtained from the VMC calculations, we find that our results agree with the experimental data at the $\lesssim 5\%$ level. As expected, two-body currents play a small role in the charge radius, providing a $\lesssim 2\%$ contribution to the total value. For the magnetic radii, we agree within the present error on the experimental data; however, future measurements of this quantity that put stronger constrains on the experimental data would provide more insight for nuclear models. Two-body corrections contribute $\lesssim 1\%$ to $\lesssim 10\%$ of the magnetic radii for nuclei that receive both isovector and isoscalar current corrections. Where the subleading isovector current is suppressed by symmetry, two-body currents play a smaller role, contributing $<1\%$ to the total result.

This letter reports on the first quantum Monte Carlo calculations of magnetic radii in $A>4$ nuclei using the $\chi$EFT approach combined with many-body methods, and contributes to the growing body of work on electric moments and point-particle radii in light systems~\cite{Hiyama:2024zfd,Caprio:2024tzt,Caprio:2025osg,Shen:2024qzi,Wolfgruber:2025rys}. These results, along with those reported in Refs.~\cite{Chambers-Wall:2024fha,Chambers-Wall:2024uhq,king:2024} for the magnetic moments, as well as longitudinal and magnetic form factors of light nuclei, combine to reveal the low-energy structure of several systems. Accurate theoretical calculations of these observables serve as necessary prerequisites for precision atomic spectroscopy~\cite{Nevo_Dinur_2019,Dickopf_2024}, and thus new interest from the theoretical community will support the interpretation of future experimental data in this area. 

{\it Theory and methods.} -- The charge radius $r_E$ of a nucleus is determined from the low $q$ behavior of the matrix element of the charge operator $\rho$, where $q$ represents the momentum transferred to the nucleus by  the external electromagnetic probe. In the limit of $q\rightarrow 0$, 
\begin{equation}
 \frac{1}{Z}\mel{JJ}{\rho(q\bfzh)}{JJ} \approx 1 - \frac{1}{6}r_E^2q^2 + \mathcal{O}(q^4)\, ,
\label{eq:ch}   
\end{equation}
where $Z$ is the proton number, and $\ket{JJ}$ is the nuclear state with angular momentum $J$ having projection $J$ along the spin-quantization axis $\bfzh$. Similarly, the magnetic radius $r_M$ is extracted from the matrix element of the electromagnetic vector current, {\bf j}, as
\begin{equation}
-i \frac{2m}{q\mu}\mel{JJ}{\mathrm{j}_y(q\bfxh)}{JJ} \approx 1 - \frac{1}{6}r_M^2q^2 + \mathcal{O}(q^4)\, ,
\label{eq:mm}
\end{equation}
where $m$ is the nucleon mass and $\mu$ is the nuclear magnetic moment of the system. The charge and magnetic radii are obtained from low-momentum polynomial fits to the matrix elements of Eqs.~(\ref{eq:ch}) and~(\ref{eq:mm}), respectively or directly from the numerical second derivative of the form factor at $q=0$.
By extracting the charge and magnetic radii directly from the electromagnetic form factors, as outlined above, this method automatically accounts for spin-orbit corrections and other nuclear effects, such as two-body correlations and currents.
We note that for $J \ge 1$ nuclei, higher multipole contributions to the form factors like $C_2$ and $M_3$ that vanish at $q=0$ can still contribute to the radii.

We computed the matrix elements using the variational Monte Carlo (VMC) method~\cite{Wiringa:1991kp,Pudliner:1997ck}, which allows for the retention of the complexity arising from many-nucleon correlations and currents. This stochastic approach to solve the many-body Schr\"odinger equation for strongly correlated nucleons has been extensively reviewed~\cite{Carlson:2014vla,Gandolfi:2020pbj}, and interested readers may consult these references for more details. 

We present results based on a single Norfolk model for the two- and three-nucleon interactions. For the magnetic form factors, there is minimal dependence on the specific Norfolk model when including currents up to next-to-next-to-next-to leading order (N3LO), and the leading order (LO) contribution also shows minimal model dependence at low-$q$~\cite{Chambers-Wall:2024uhq,Chambers-Wall:2024fha}. Furthermore, for values of $q \leq 2.5$ fm, the longitudinal form factors are consistent across models~\cite{king:2024}. Because the charge and magnetic radii are constrained by the low-$q$ behavior of the corresponding form factors, these findings reinforce our choice to present results for one model. 

Specifically, the nuclear Hamiltonian is taken to consist of the single nucleon kinetic terms supplemented by one of the Norfolk two- and three nucleon interactions. We adopt the Norfolk IIb$^{\star}$  potential~\cite{Piarulli:2014bda,Piarulli:2016vel,Piarulli:2017dwd,Baroni:2018fdn,Piarulli:2019cqu}, that describes long- and intermediate parts of the interactions via one- and multiple-pion exchange contributions, while encoding the short-range dynamics in contact terms. In this model, the two-body low energy constants (LECs) are constrained using 3695 nucleon-nucleon scattering data up to 200 MeV in the lab frame with a $\chi^2$ per datum $\sim 1.37$, and is characterized by two cutoff parameters, $[R_L,R_S]=[1.0~{\rm fm},0.7~{\rm fm}]$~\cite{Piarulli:2016vel}. These appear in the regularization functions designed to eliminate singularities at very small inter-particle distances, and in the Gaussian smearing of the delta functions associated with two-nucleon contact terms, respectively. The three-nucleon force involves two LECs that are constrained by the trinucleon binding energies and the $^3$H Gamow-Teller matrix element~\cite{Baroni:2018fdn}. 

The coupling of the many-body system to the external electromagnetic fields is implemented via one- and two-nucleon charge and current operators from $\chi$EFT~\cite{Park:1995pn,WALZL200137,PHILLIPS200312,Phillips_2007,Kolling:2009iq,Kolling:2011mt,Krebs:2016rqz,Krebs:2019aka,Pastore:2008ui,Pastore:2009is,Pastore:2011ip,Piarulli:2012bn,Schiavilla:2018udt,Gnech:2022vwr}. Specifically, we adopt the operators derived in Refs.~\cite{Pastore:2008ui,Pastore:2009is,Pastore:2011ip,Piarulli:2012bn,Schiavilla:2018udt} within the same theoretical framework as the Norfolk interactions. The electromagnetic charge, $\rho^{\rm LO}$, and current, ${\bf j}^{\rm LO}$, operators at LO in the chiral expansion are derived from the non-relativistic reduction of the covariant single-nucleon current, which leads to the standard Impulse Approximation expressions given by~\cite{Carlson:1997qn},
\begin{eqnarray}
\label{eq:lo}
\rho^{\rm LO} &=&  \epsilon_i(q^2) \,{\rm e}^{i{\bf q}\cdot{\bf r}_i} \ , \\
{\bf j}^{\rm LO}&=&\frac{\epsilon_i(q^2) }{2\, m}\,
\left[{\bf p}_i\,\, ,\, \, {\rm e}^{i{\bf q}\cdot{\bf r}_i}  \right]_+ 
+i\,\frac{\mu_{i}(q^2) }{2\, m} \,\, {\rm e}^{i{\bf q}\cdot{\bf r}_i}\,\, {\bm \sigma}_i\times {\bf q }\ , \nonumber
\end{eqnarray}
where ${\bf p}_i\,$=$\, -i\, {\bm \nabla}_i$, ${\bf r}_i$ and ${\bm \sigma}_i$ are the single-nucleon coordinate and spin, respectively, $m$ is the average nucleon mass, and $e$ is the unit of charge. The brackets $[ \ldots,\ldots]_{+}$ denote the anti-commutator. The charge and magnetic distributions  of the nucleons are encoded in the $\epsilon_i$ and $\mu_{i}$ operators defined as
\begin{eqnarray}
\epsilon_{i}(q^2) &=& \frac{G_E^S(q^2)+G_E^V(q^2)\, \tau_{i,z}}{2}\ , \nonumber \\ 
 \mu_{i}(q^2) &=&\frac{G_M^S(q^2)+G_M^V(q^2)\, \tau_{i,z}}{2} \ ,
\end{eqnarray}
where $G^{S/V}_E$ and $G^{S/V}_M$ denote the isoscalar/isovector combinations of the proton and neutron electric ($E$) and magnetic ($M$) form factors, normalized as $G^S_E(0)=G^V_E(0)=1$, $G^S_M(0)=0.880
\, \mu_N$, and $G^V_M(0)=4.706\, \mu_N$, where $\mu_N$ is the nuclear magneton. It is worth noting that we use dipole form factors extracted from experimental electron scattering data on the proton and deuteron~\cite{Kelly:2004}; however, on-going and planned experimental efforts aim to provide clarity on the question of the proton radius~\cite{Suda:2022hsm,PRad:2020oor,A1:2022wzx,Cline:2021ehf,Friedrich:2024ylw}, and as a result, more data will be available on structure of the nucleons. 
In the future, it may be worth investigating the sensitivity of the nuclear form factors to the nucleon structure by, {\it e.g.}, exploring the effect of using different parameterizations of the electromagnetic nucleonic form factors, integrating new insights from these experiments and from the recent theoretical advances from $\chi$EFT~\cite{Filin:2019eoe,Filin:2020tcs}. 

We include sub-leading terms up to N3LO in the charge and current operators.  
For the electromagnetic vector current, two-nucleon terms of one-pion range already emerge at next-to leading order (NLO) in the chiral expansion. These terms are purely isovector and are crucial for explaining the magnetic structure of nuclei~\cite{Carlson:1997qn,Pastore:2012rp,Chambers-Wall:2024fha,Chambers-Wall:2024uhq}. At next-to-next-to leading order (N2LO), two contributions appear: the first is a one-body term that arises from incorporating relativistic corrections to the LO operators of Eqs.~(\ref{eq:lo}); the second is a two-body term generated by the excitation of a $\Delta$ that emits a pion reabsorbed by the second nucleon. At N3LO, there are contributions of both one- and two-pion range, along with contact-like currents encoding short-range dynamics. The electromagnetic current at N3LO involves five unknown low-energy constant that are constrained to reproduce the magnetic moments of the deuteron and the trinucleon systems, as well as backward-angle deuteron electron-disintegration data at threshold~\cite{Gnech:2022vwr}. 

The expansion of the charge operator is rather different from that of the vector current. The first correction appears at N2LO, where only the one-body relativistic correction to the LO operator appears. Here, two-nucleon contributions arise at N3LO and involve the exchange of a pion with no additional unknown low-energy constants. These terms are all proportional to $1/m$, where $m$ is the nucleon mass, and can be regarded as small kinematic corrections. In the study of Ref.~\cite{King:2024zbv}, they are found to provide a minimal contribution to the calculated charge form factor at the low momentum transfers needed to extract the charge radius.  

{\it Results.} -- 

\begin{table}[tbh]
\begin{center}
\begin{tabular} {c | c | c | c }\hline \hline 
 & $r^{\rm LO}_E$ (fm) &  $r^{\rm Tot}_E$ (fm) &Expt. (fm) \\ \hline
\nuc{3}{H}$(\frac{1}{2}^+;\frac{1}{2})$  &1.69(1) &1.72(1) &1.755(86)~\cite{Amroun:1994qj}  \\  
\nuc{3}{He}$(\frac{1}{2}^+;\frac{1}{2})$ &1.90(1)  &1.92(1)  &1.9506(14)~\cite{Shiner:1995zz}\\  
\nuc{4}{He}$(0^+;0)$ & 1.64(1) &1.67(1) &1.67824(83)~\cite{Krauth:2021foz} \\ 
\nuc{6}{He}$(0^+;1)$ &2.07(1) & 2.07(1)  & 2.059(8)~\cite{Lu:2013ena} \\
\nuc{6}{Li}$(1^+;0)$                     &2.58(3) &2.60(3) &2.589(39)~\cite{Nortershauser:2011zz}\\
\nuc{7}{Li}$(\frac{3}{2}^-;\frac{1}{2})$ &2.35(2) &2.37(2) &2.444(42)~\cite{Nortershauser:2011zz}\\
\nuc{7}{Be}$(\frac{3}{2}^-;\frac{1}{2})$ &2.53(2) &2.55(3) &2.647(17)~\cite{Nortershauser:2008vp}\\
\nuc{8}{He}$(0^+;2)$ &1.97(1) &1.91(9)   & 1.958(16)~\cite{Lu:2013ena} \\ 
\nuc{8}{Li}$(2^+;1)$                     &2.32(2) &2.32(3) &2.339(44)~\cite{Sick:2008zza} \\
\nuc{8}{Be}$(0^+;0)$ &2.53(2) &2.55(2) &--\\
\nuc{8}{B}$(2^+;1)$                      &2.63(3) &2.67(4) &--\\
\nuc{8}{C}$(0^+;2)^{\dagger}$ &2.88(4) &2.91(5)  &-- \\ 
\nuc{9}{Li}$(\frac{3}{2}^-;\frac{3}{2})$ &2.25(2) &2.25(4) &2.245(46)~\cite{Nortershauser:2011zz} \\ 
\nuc{9}{Be}$(\frac{3}{2}^-;\frac{1}{2})$ &2.45(2) &2.46(2) &2.519(12)~\cite{Jansen:1972iui}\\ 
\nuc{9}{B}$(\frac{3}{2}^-;\frac{1}{2})^{\dagger}$  &2.55(2)  &2.59(3) &-- \\ 
\nuc{9}{C}$(\frac{3}{2}^-;\frac{3}{2})^{\dagger}$  &2.67(3) &2.70(4) &--\\
\nuc{10}{B}$(3^+;0)$                     &2.45(2)  &2.47(2) &2.58(7)~\cite{Cichocki:1995zz} \\
\hline\hline
\end{tabular}
\end{center}
\caption{LO ($r^{\rm LO}_E$) and N3LO ($r^{\rm Tot}_E$) charge radii computed using VMC retaining with the Norfolk model IIb$^{\star}$. Experimental data are presented in the third column. The numbers in parentheses denote the theoretical uncertainty for the VMC results or the experimental error for the data.}
\label{tab:vmc.chg.radii}
\end{table}
\begin{table}[tbh]
\begin{center}
\begin{tabular} {c | c | c | c }\hline \hline 
 & $r^{\rm LO}_M$ (fm)  & $r^{\rm Tot}_M$ (fm) &Expt (fm) \\ \hline
\nuc{3}{H}$(\frac{1}{2}^+;\frac{1}{2})$  &1.88(2) &1.82(1)  &1.840(181)~\cite{Amroun:1994qj} \\  
\nuc{3}{He}$(\frac{1}{2}^+;\frac{1}{2})$ &2.02(3) &1.92(2)  &1.965(153)~\cite{Amroun:1994qj}  \\  
\nuc{6}{Li}$(1^+;0)$                     &3.32(10) &3.32(10)  &-- \\
\nuc{7}{Li}$(\frac{3}{2}^-;\frac{1}{2})$ & 2.89(7) &2.99(29) &--  \\
\nuc{7}{Be}$(\frac{3}{2}^-;\frac{1}{2})$ & 3.42(11) &3.37(31) &--  \\             
\nuc{8}{Li}$(2^+;1)$                     & 2.22(2) &2.31(1) &-- \\
\nuc{8}{B}$(2^+;1)$                      & 3.04(4) &3.25(2) &-- \\
\nuc{9}{Li}$(\frac{3}{2}^-;\frac{3}{2})$ & 2.80(7) &2.87(31) &-- \\ 
\nuc{9}{Be}$(\frac{3}{2}^-;\frac{1}{2})$ & 3.34(7) &3.28(7) &-- \\ 
\nuc{9}{B}$(\frac{3}{2}^-;\frac{1}{2})^{\dagger}$  & 2.80(9) &2.82(12) &-- \\ 
\nuc{9}{C}$(\frac{3}{2}^-;\frac{3}{2})^{\dagger}$  & 3.34(7) &3.14(30) &--  \\
\nuc{10}{B}$(3^+;0)$                     & 2.33(2) &2.33(2) &--  \\
\hline\hline
\end{tabular}
\end{center}
\caption{LO ($r^{\rm LO}_M$) and N3LO ($r^{\rm Tot}_M$) magnetic radii computed using VMC retaining with the Norfolk model IIb$^{\star}$. The numbers in parentheses denote the theoretical uncertainty for the VMC results or the experimental error for the data.}
\label{tab:vmc.mag.radii}
\end{table}

We present electromagnetic radii computed in VMC-- using both the charge and current operators at LO and retaining higher-order corrections in the $\chi$EFT expansion up to N3LO-- in Tables~\ref{tab:vmc.chg.radii} and~\ref{tab:vmc.mag.radii}. Results for $r_E$ in Table~\ref{tab:vmc.chg.radii} are compared to experimental data where available. For the results of $r_M$ in Table~\ref{tab:vmc.mag.radii}, there are presently only data for nuclei with mass $A\leq3$ that we take from Ref.~\cite{Amroun:1994qj}. Finally, nuclei marked with a dagger have their VMC wave functions generated simply by swapping protons for neutrons in the optimized wave function of the isobar analogue. Should data become available for these systems, a more sophisticated treatment of the wave function, accounting for charge symmetry breaking effects, and a detailed analysis of model dependencies  may be necessary. Additionally, we note that the results presented are an average of the fitting and derivative extraction procedures, and that the error reflects only the variation in these two approaches. Thus, our results neglect model dependencies and variations arising  from the use of different parametrizations of the nucleons' electromagnetic form factors. 

\begin{figure}
\includegraphics[width=0.4\textwidth]{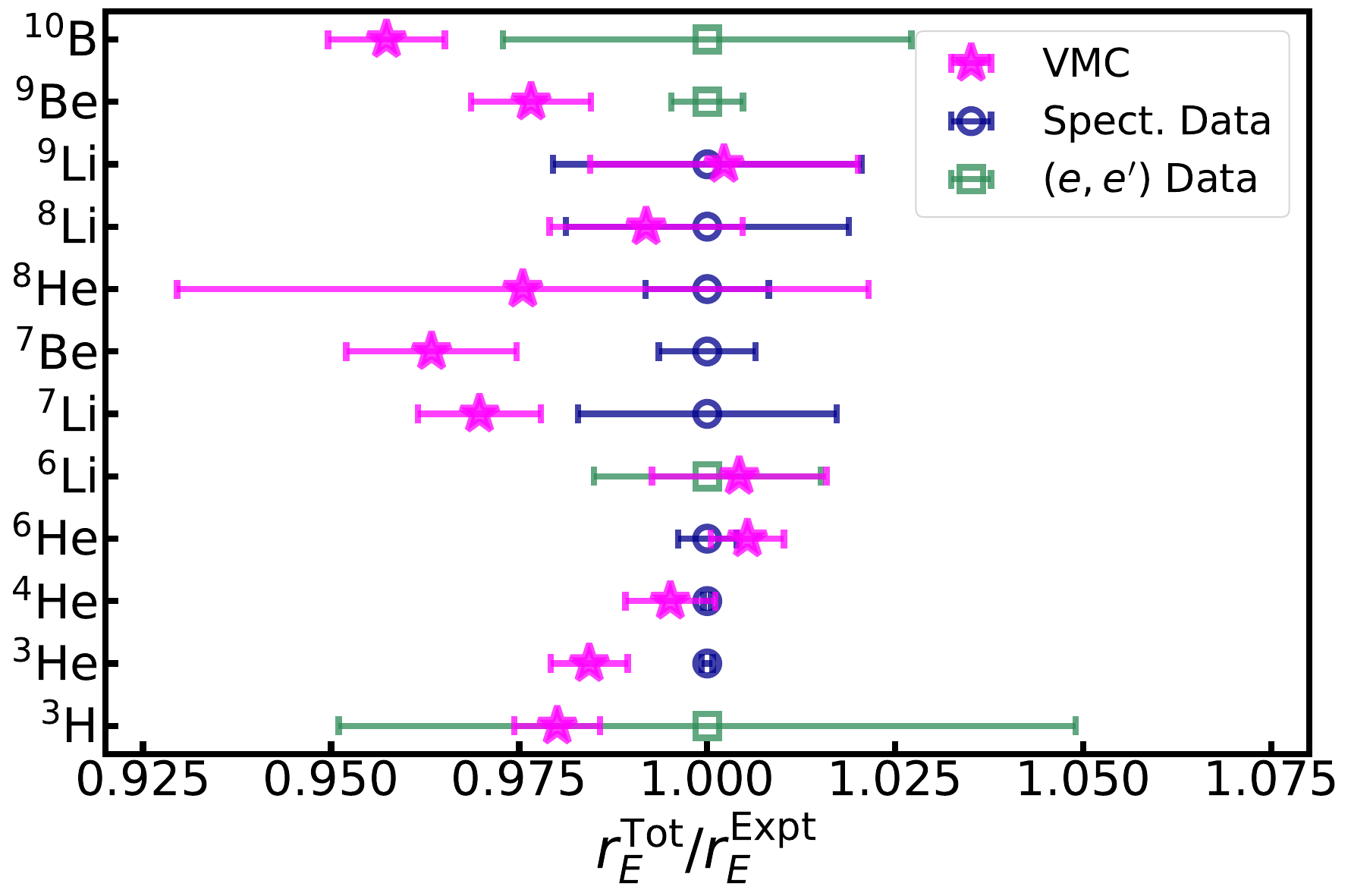}
    \caption{The ratio of charge radii $r^{\rm Tot}_E$ computed with VMC (mageta stars) relative to experimental data from atomic spectroscopy (open blue circles) and electron scattering (open green squares) measurements for various nuclei. The magenta (blue and green) bars represent theoretical (experimental) uncertainty. The experimental data are from Refs.~\cite{Amroun:1994qj,Shiner:1995zz,Sick:2008zza,Lu:2013ena,Nortershauser:2008vp,Nortershauser:2011zz,Cichocki:1995zz}.}
\label{fig:rad.ratio}
\end{figure}

\begin{figure}
\includegraphics[width=0.4\textwidth]{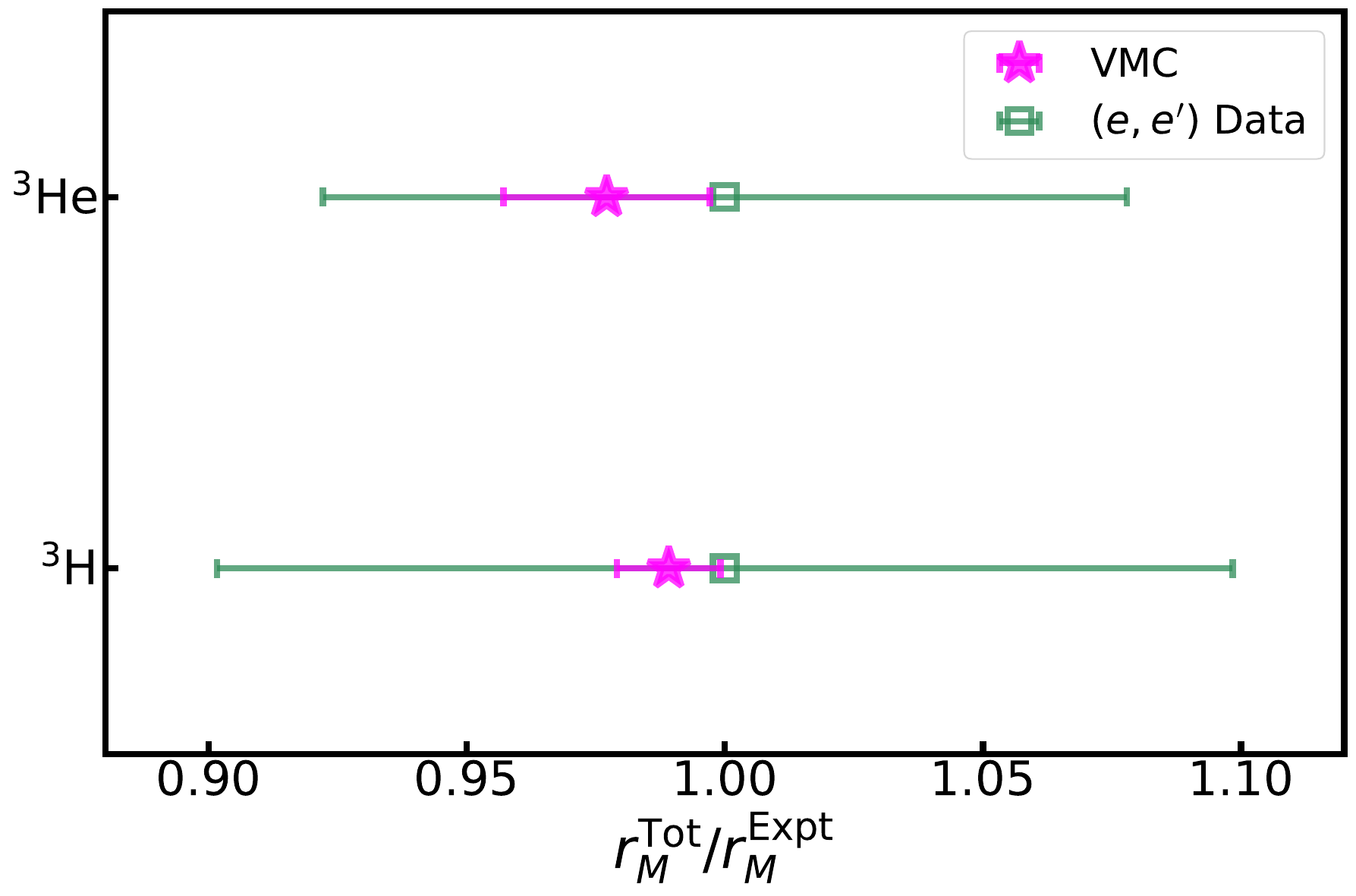}
    \caption{The same as Fig.~\ref{fig:rad.ratio}, but for $r_M$. The experimental data are from Refs.~\cite{Amroun:1994qj}.}
\label{fig:rad.ratio.m}
\end{figure}

For the charge radii, we present both the leading order $r^{\rm LO}_E$ results and the results containing operators up to N3LO in the chiral expansion $r^{\rm Tot}_E$. The two-body current contribution to this quantity is small, making up only $\lesssim 2\%$ of the total value across the nuclei studied, and it is generally negligible compared to the error from different extraction procedures.  In general $r_E$ increases slightly going from LO to Tot.  A notable exception is $^8$He, which has a large negative one-body spin-orbit contribution appearing at N2LO which makes the Tot value less than the LO by 0.06 fm.  Other cases where the one-body spin-orbit is negative and big enough to prevent growth going from LO to Tot are $^6$He and $^{8,9}$Li.

To visualize the comparison of $r^{\rm Tot}_E$ with data, Figure~\ref{fig:rad.ratio} shows the VMC values as a ratio to experiment. For the central values, we find agreement  within ${\sim}4\%$ in all cases. It is interesting to note that the results which are not compatible with the data tend to under estimate the experimental values. It is also worth noting that, although we adequately describe the charge form factors of $^9$Be~\cite{Jansen:1972iui} and $^{10}$B~\cite{Cichocki:1995zz}, we do not agree with the charge radii obtained from extrapolating fits of that same form factor data to low $q$. As such, lower $q$ data for the charge form factors-- which may be feasible with advances in low-energy electron accelerator facilities~\cite{Suda:2022hsm,Bernauer:2022cqe} -- would help to determine whether the many-body calculation truly does not describe nature, or if the discrepancy arises due to the extrapolation of the current data. In fact, recent studies on moments of nuclear charge distributions of the trinucleon systems~\cite{Hiyama:2024zfd} have also highlighted the need for accurate electron-scattering experimental data at small $q$ for a reliable extraction of nuclear moments. Furthermore, the experimental data reported in Table~\ref{tab:vmc.chg.radii} comes from electron scattering or laser spectroscopy, the extraction of which generally relies on some nuclear or atomic model, respectively. Novel techniques to measure electron scattering from on-line produced radioisotopes~\cite{Tsukada:2023,Suda:2022hsm}, may provide yet another way to benchmark the predicted charge radii of exotic nuclei to validate nuclear theory, and would complement laser spectroscopy efforts~\cite{Minamisono:2013,Ohayon:2023hze}.

In the case of the magnetic radii, we once again present the leading order $r^{\rm LO}_M$ results and the value retaining currents through N3LO $r^{\rm Tot}_M$. Two-body currents play a small role in the isoscalar nuclei, contributing $<1\%$ of the total radius. Where isovector currents are not suppressed by symmetry, higher-order currents play a larger role that is, however, more modest than for magnetic moments. Although magnetic moments can have up to $33\%$ of their contribution coming from two-body currents~\cite{Pastore:2012rp,Chambers-Wall:2024uhq}, the radii receive modest contributions of $\lesssim 10\%$. This behavior can be understood from the definition of magnetic radii given in Eq.~(\ref{eq:mm}). The NLO correction is to $r_M^2$, and thus the two-body current effect is smaller on $r_M$ itself. In this case the only available experimental data are for $^3$H and $^3$He, which are extracted from the magnetic form factors in Ref.~\cite{Amroun:1994qj} and have relatively large errors. Thus, new measurements of magnetic properties of nuclei using atomic spectroscopy could allow precise tests of both nuclear and atomic models.

{\it Acknowledgements} --  S.P. would like to thank E.~Hiyama and N.~Yamanaka for their hospitality at RIKEN and for insightful discussions. This work is supported by the US Department of Energy under Contracts No. DE-SC0021027 (G.~B.~K., G.~C.-W., and S.~P.), DE-AC02-06CH11357 (R.B.W.), DE-AC05-06OR23177 (A.G.), a 2021 Early Career Award number DE-SC0022002 (M.~P.), the FRIB Theory Alliance award DE-SC0013617 (M.~P.), and the NUCLEI SciDAC program (S.P., M.P., and R.B.W.). G.~B.~K. would like to acknowledge support from the U.S. DOE NNSA Stewardship Science Graduate Fellowship under Cooperative Agreement DE-NA0003960, and from the Laboratory Directed Research and Development program of Los Alamos National Laboratory under project number 20240742PRD1. G.~C.-W. acknowledges support from the NSF Graduate Research Fellowship Program under Grant No. DGE-213989. We thank the Nuclear Theory for New Physics Topical Collaboration, supported by the U.S.~Department of Energy under contract DE-SC0023663, for fostering dynamic collaborations.
 A.G. acknowledges the direct support of Nuclear Theory for New Physics Topical collaboration.

The many-body calculations were performed on the parallel computers of the Laboratory Computing Resource Center, Argonne National Laboratory, the computers of the Argonne Leadership Computing Facility (ALCF) via the INCITE grant ``Ab-initio nuclear structure and nuclear reactions'', the 2019/2020 ALCC grant ``Low Energy Neutrino-Nucleus interactions'' for the project NNInteractions, the 2020/2021 ALCC grant ``Chiral Nuclear Interactions from Nuclei to Nucleonic Matter'' for the project ChiralNuc, the 2021/2022 ALCC grant ``Quantum Monte Carlo Calculations of Nuclei up to $^{16}{\rm O}$ and Neutron Matter" for the project \mbox{QMCNuc}, and by the National Energy Research
Scientific Computing Center, a DOE Office of Science User Facility
supported by the Office of Science of the U.S. Department of Energy
under Contract No. DE-AC02-05CH11231 using NERSC award
NP-ERCAP0027147.

\bibliography{biblio}

\begin{thebibliography}{57}%
\makeatletter
\providecommand \@ifxundefined [1]{%
 \@ifx{#1\undefined}
}%
\providecommand \@ifnum [1]{%
 \ifnum #1\expandafter \@firstoftwo
 \else \expandafter \@secondoftwo
 \fi
}%
\providecommand \@ifx [1]{%
 \ifx #1\expandafter \@firstoftwo
 \else \expandafter \@secondoftwo
 \fi
}%
\providecommand \natexlab [1]{#1}%
\providecommand \enquote  [1]{``#1''}%
\providecommand \bibnamefont  [1]{#1}%
\providecommand \bibfnamefont [1]{#1}%
\providecommand \citenamefont [1]{#1}%
\providecommand \href@noop [0]{\@secondoftwo}%
\providecommand \href [0]{\begingroup \@sanitize@url \@href}%
\providecommand \@href[1]{\@@startlink{#1}\@@href}%
\providecommand \@@href[1]{\endgroup#1\@@endlink}%
\providecommand \@sanitize@url [0]{\catcode `\\12\catcode `\$12\catcode `\&12\catcode `\#12\catcode `\^12\catcode `\_12\catcode `\%12\relax}%
\providecommand \@@startlink[1]{}%
\providecommand \@@endlink[0]{}%
\providecommand \url  [0]{\begingroup\@sanitize@url \@url }%
\providecommand \@url [1]{\endgroup\@href {#1}{\urlprefix }}%
\providecommand \urlprefix  [0]{URL }%
\providecommand \Eprint [0]{\href }%
\providecommand \doibase [0]{http://dx.doi.org/}%
\providecommand \selectlanguage [0]{\@gobble}%
\providecommand \bibinfo  [0]{\@secondoftwo}%
\providecommand \bibfield  [0]{\@secondoftwo}%
\providecommand \translation [1]{[#1]}%
\providecommand \BibitemOpen [0]{}%
\providecommand \bibitemStop [0]{}%
\providecommand \bibitemNoStop [0]{.\EOS\space}%
\providecommand \EOS [0]{\spacefactor3000\relax}%
\providecommand \BibitemShut  [1]{\csname bibitem#1\endcsname}%
\let\auto@bib@innerbib\@empty
\bibitem [{\citenamefont {Chambers-Wall}\ \emph {et~al.}(2024{\natexlab{a}})\citenamefont {Chambers-Wall}, \citenamefont {Gnech}, \citenamefont {King}, \citenamefont {Pastore}, \citenamefont {Piarulli}, \citenamefont {Schiavilla},\ and\ \citenamefont {Wiringa}}]{Chambers-Wall:2024fha}%
  \BibitemOpen
  \bibfield  {author} {\bibinfo {author} {\bibfnamefont {G.}~\bibnamefont {Chambers-Wall}}, \bibinfo {author} {\bibfnamefont {A.}~\bibnamefont {Gnech}}, \bibinfo {author} {\bibfnamefont {G.~B.}\ \bibnamefont {King}}, \bibinfo {author} {\bibfnamefont {S.}~\bibnamefont {Pastore}}, \bibinfo {author} {\bibfnamefont {M.}~\bibnamefont {Piarulli}}, \bibinfo {author} {\bibfnamefont {R.}~\bibnamefont {Schiavilla}}, \ and\ \bibinfo {author} {\bibfnamefont {R.~B.}\ \bibnamefont {Wiringa}},\ }\href@noop {} {\  (\bibinfo {year} {2024}{\natexlab{a}})},\ \Eprint {http://arxiv.org/abs/2407.03487} {arXiv:2407.03487 [nucl-th]} \BibitemShut {NoStop}%
\bibitem [{\citenamefont {Chambers-Wall}\ \emph {et~al.}(2024{\natexlab{b}})\citenamefont {Chambers-Wall}, \citenamefont {Gnech}, \citenamefont {King}, \citenamefont {Pastore}, \citenamefont {Piarulli}, \citenamefont {Schiavilla},\ and\ \citenamefont {Wiringa}}]{Chambers-Wall:2024uhq}%
  \BibitemOpen
  \bibfield  {author} {\bibinfo {author} {\bibfnamefont {G.}~\bibnamefont {Chambers-Wall}}, \bibinfo {author} {\bibfnamefont {A.}~\bibnamefont {Gnech}}, \bibinfo {author} {\bibfnamefont {G.~B.}\ \bibnamefont {King}}, \bibinfo {author} {\bibfnamefont {S.}~\bibnamefont {Pastore}}, \bibinfo {author} {\bibfnamefont {M.}~\bibnamefont {Piarulli}}, \bibinfo {author} {\bibfnamefont {R.}~\bibnamefont {Schiavilla}}, \ and\ \bibinfo {author} {\bibfnamefont {R.~B.}\ \bibnamefont {Wiringa}},\ }\href@noop {} {\  (\bibinfo {year} {2024}{\natexlab{b}})},\ \Eprint {http://arxiv.org/abs/2407.04744} {arXiv:2407.04744 [nucl-th]} \BibitemShut {NoStop}%
\bibitem [{\citenamefont {King}\ \emph {et~al.}()\citenamefont {King}, \citenamefont {Chambers-Wall}, \citenamefont {Gnech}, \citenamefont {Pastore}, \citenamefont {Piarulli}, \citenamefont {Schiavilla},\ and\ \citenamefont {Wiringa}}]{king:2024}%
  \BibitemOpen
  \bibfield  {author} {\bibinfo {author} {\bibfnamefont {G.~B.}\ \bibnamefont {King}}, \bibinfo {author} {\bibfnamefont {G.}~\bibnamefont {Chambers-Wall}}, \bibinfo {author} {\bibfnamefont {A.}~\bibnamefont {Gnech}}, \bibinfo {author} {\bibfnamefont {S.}~\bibnamefont {Pastore}}, \bibinfo {author} {\bibfnamefont {M.}~\bibnamefont {Piarulli}}, \bibinfo {author} {\bibfnamefont {R.}~\bibnamefont {Schiavilla}}, \ and\ \bibinfo {author} {\bibfnamefont {R.~B.}\ \bibnamefont {Wiringa}},\ }\href@noop {} {\enquote {\bibinfo {title} {{Longitudinal form factors of $A\,\leq\,10$ nuclei in a chiral effective field theory approach}},}\ }\bibinfo {note} {Submitted to {\it Physical Review C}}\BibitemShut {NoStop}%
\bibitem [{\citenamefont {Piarulli}\ \emph {et~al.}(2015)\citenamefont {Piarulli}, \citenamefont {Girlanda}, \citenamefont {Schiavilla}, \citenamefont {Navarro~Pérez}, \citenamefont {Amaro},\ and\ \citenamefont {Ruiz~Arriola}}]{Piarulli:2014bda}%
  \BibitemOpen
  \bibfield  {author} {\bibinfo {author} {\bibfnamefont {M.}~\bibnamefont {Piarulli}}, \bibinfo {author} {\bibfnamefont {L.}~\bibnamefont {Girlanda}}, \bibinfo {author} {\bibfnamefont {R.}~\bibnamefont {Schiavilla}}, \bibinfo {author} {\bibfnamefont {R.}~\bibnamefont {Navarro~Pérez}}, \bibinfo {author} {\bibfnamefont {J.~E.}\ \bibnamefont {Amaro}}, \ and\ \bibinfo {author} {\bibfnamefont {E.}~\bibnamefont {Ruiz~Arriola}},\ }\href {\doibase 10.1103/PhysRevC.91.024003} {\bibfield  {journal} {\bibinfo  {journal} {Phys. Rev.}\ }\textbf {\bibinfo {volume} {C91}},\ \bibinfo {pages} {024003} (\bibinfo {year} {2015})},\ \Eprint {http://arxiv.org/abs/1412.6446} {arXiv:1412.6446 [nucl-th]} \BibitemShut {NoStop}%
\bibitem [{\citenamefont {Piarulli}\ \emph {et~al.}(2016)\citenamefont {Piarulli}, \citenamefont {Girlanda}, \citenamefont {Schiavilla}, \citenamefont {Kievsky}, \citenamefont {Lovato}, \citenamefont {Marcucci}, \citenamefont {Pieper}, \citenamefont {Viviani},\ and\ \citenamefont {Wiringa}}]{Piarulli:2016vel}%
  \BibitemOpen
  \bibfield  {author} {\bibinfo {author} {\bibfnamefont {M.}~\bibnamefont {Piarulli}}, \bibinfo {author} {\bibfnamefont {L.}~\bibnamefont {Girlanda}}, \bibinfo {author} {\bibfnamefont {R.}~\bibnamefont {Schiavilla}}, \bibinfo {author} {\bibfnamefont {A.}~\bibnamefont {Kievsky}}, \bibinfo {author} {\bibfnamefont {A.}~\bibnamefont {Lovato}}, \bibinfo {author} {\bibfnamefont {L.~E.}\ \bibnamefont {Marcucci}}, \bibinfo {author} {\bibfnamefont {S.~C.}\ \bibnamefont {Pieper}}, \bibinfo {author} {\bibfnamefont {M.}~\bibnamefont {Viviani}}, \ and\ \bibinfo {author} {\bibfnamefont {R.~B.}\ \bibnamefont {Wiringa}},\ }\href {\doibase 10.1103/PhysRevC.94.054007} {\bibfield  {journal} {\bibinfo  {journal} {Phys. Rev.}\ }\textbf {\bibinfo {volume} {C94}},\ \bibinfo {pages} {054007} (\bibinfo {year} {2016})},\ \Eprint {http://arxiv.org/abs/1606.06335} {arXiv:1606.06335 [nucl-th]} \BibitemShut {NoStop}%
\bibitem [{\citenamefont {Piarulli}\ \emph {et~al.}(2018)\citenamefont {Piarulli} \emph {et~al.}}]{Piarulli:2017dwd}%
  \BibitemOpen
  \bibfield  {author} {\bibinfo {author} {\bibfnamefont {M.}~\bibnamefont {Piarulli}} \emph {et~al.},\ }\href {\doibase 10.1103/PhysRevLett.120.052503} {\bibfield  {journal} {\bibinfo  {journal} {Phys. Rev. Lett.}\ }\textbf {\bibinfo {volume} {120}},\ \bibinfo {pages} {052503} (\bibinfo {year} {2018})},\ \Eprint {http://arxiv.org/abs/1707.02883} {arXiv:1707.02883 [nucl-th]} \BibitemShut {NoStop}%
\bibitem [{\citenamefont {Baroni}\ \emph {et~al.}(2018)\citenamefont {Baroni} \emph {et~al.}}]{Baroni:2018fdn}%
  \BibitemOpen
  \bibfield  {author} {\bibinfo {author} {\bibfnamefont {A.}~\bibnamefont {Baroni}} \emph {et~al.},\ }\href {\doibase 10.1103/PhysRevC.98.044003} {\bibfield  {journal} {\bibinfo  {journal} {Phys. Rev.}\ }\textbf {\bibinfo {volume} {C98}},\ \bibinfo {pages} {044003} (\bibinfo {year} {2018})},\ \Eprint {http://arxiv.org/abs/1806.10245} {arXiv:1806.10245 [nucl-th]} \BibitemShut {NoStop}%
\bibitem [{\citenamefont {Piarulli}\ and\ \citenamefont {Tews}(2020)}]{Piarulli:2019cqu}%
  \BibitemOpen
  \bibfield  {author} {\bibinfo {author} {\bibfnamefont {M.}~\bibnamefont {Piarulli}}\ and\ \bibinfo {author} {\bibfnamefont {I.}~\bibnamefont {Tews}},\ }\href {\doibase 10.3389/fphy.2019.00245} {\bibfield  {journal} {\bibinfo  {journal} {Front. in Phys.}\ }\textbf {\bibinfo {volume} {7}},\ \bibinfo {pages} {245} (\bibinfo {year} {2020})},\ \Eprint {http://arxiv.org/abs/2002.00032} {arXiv:2002.00032 [nucl-th]} \BibitemShut {NoStop}%
\bibitem [{\citenamefont {Pastore}\ \emph {et~al.}(2008)\citenamefont {Pastore}, \citenamefont {Schiavilla},\ and\ \citenamefont {Goity}}]{Pastore:2008ui}%
  \BibitemOpen
  \bibfield  {author} {\bibinfo {author} {\bibfnamefont {S.}~\bibnamefont {Pastore}}, \bibinfo {author} {\bibfnamefont {R.}~\bibnamefont {Schiavilla}}, \ and\ \bibinfo {author} {\bibfnamefont {J.~L.}\ \bibnamefont {Goity}},\ }\href {\doibase 10.1103/PhysRevC.78.064002} {\bibfield  {journal} {\bibinfo  {journal} {Phys. Rev.}\ }\textbf {\bibinfo {volume} {C78}},\ \bibinfo {pages} {064002} (\bibinfo {year} {2008})},\ \Eprint {http://arxiv.org/abs/0810.1941} {arXiv:0810.1941 [nucl-th]} \BibitemShut {NoStop}%
\bibitem [{\citenamefont {Pastore}\ \emph {et~al.}(2009)\citenamefont {Pastore}, \citenamefont {Girlanda}, \citenamefont {Schiavilla}, \citenamefont {Viviani},\ and\ \citenamefont {Wiringa}}]{Pastore:2009is}%
  \BibitemOpen
  \bibfield  {author} {\bibinfo {author} {\bibfnamefont {S.}~\bibnamefont {Pastore}}, \bibinfo {author} {\bibfnamefont {L.}~\bibnamefont {Girlanda}}, \bibinfo {author} {\bibfnamefont {R.}~\bibnamefont {Schiavilla}}, \bibinfo {author} {\bibfnamefont {M.}~\bibnamefont {Viviani}}, \ and\ \bibinfo {author} {\bibfnamefont {R.~B.}\ \bibnamefont {Wiringa}},\ }\href {\doibase 10.1103/PhysRevC.80.034004} {\bibfield  {journal} {\bibinfo  {journal} {Phys. Rev.}\ }\textbf {\bibinfo {volume} {C80}},\ \bibinfo {pages} {034004} (\bibinfo {year} {2009})},\ \Eprint {http://arxiv.org/abs/0906.1800} {arXiv:0906.1800 [nucl-th]} \BibitemShut {NoStop}%
\bibitem [{\citenamefont {Pastore}\ \emph {et~al.}(2011)\citenamefont {Pastore}, \citenamefont {Girlanda}, \citenamefont {Schiavilla},\ and\ \citenamefont {Viviani}}]{Pastore:2011ip}%
  \BibitemOpen
  \bibfield  {author} {\bibinfo {author} {\bibfnamefont {S.}~\bibnamefont {Pastore}}, \bibinfo {author} {\bibfnamefont {L.}~\bibnamefont {Girlanda}}, \bibinfo {author} {\bibfnamefont {R.}~\bibnamefont {Schiavilla}}, \ and\ \bibinfo {author} {\bibfnamefont {M.}~\bibnamefont {Viviani}},\ }\href {\doibase 10.1103/PhysRevC.84.024001} {\bibfield  {journal} {\bibinfo  {journal} {Phys. Rev.}\ }\textbf {\bibinfo {volume} {C84}},\ \bibinfo {pages} {024001} (\bibinfo {year} {2011})},\ \Eprint {http://arxiv.org/abs/1106.4539} {arXiv:1106.4539 [nucl-th]} \BibitemShut {NoStop}%
\bibitem [{\citenamefont {Piarulli}\ \emph {et~al.}(2013)\citenamefont {Piarulli}, \citenamefont {Girlanda}, \citenamefont {Marcucci}, \citenamefont {Pastore}, \citenamefont {Schiavilla},\ and\ \citenamefont {Viviani}}]{Piarulli:2012bn}%
  \BibitemOpen
  \bibfield  {author} {\bibinfo {author} {\bibfnamefont {M.}~\bibnamefont {Piarulli}}, \bibinfo {author} {\bibfnamefont {L.}~\bibnamefont {Girlanda}}, \bibinfo {author} {\bibfnamefont {L.~E.}\ \bibnamefont {Marcucci}}, \bibinfo {author} {\bibfnamefont {S.}~\bibnamefont {Pastore}}, \bibinfo {author} {\bibfnamefont {R.}~\bibnamefont {Schiavilla}}, \ and\ \bibinfo {author} {\bibfnamefont {M.}~\bibnamefont {Viviani}},\ }\href {\doibase 10.1103/PhysRevC.87.014006} {\bibfield  {journal} {\bibinfo  {journal} {Phys. Rev.}\ }\textbf {\bibinfo {volume} {C87}},\ \bibinfo {pages} {014006} (\bibinfo {year} {2013})},\ \Eprint {http://arxiv.org/abs/1212.1105} {arXiv:1212.1105 [nucl-th]} \BibitemShut {NoStop}%
\bibitem [{\citenamefont {Schiavilla}\ \emph {et~al.}(2019)\citenamefont {Schiavilla}, \citenamefont {Baroni}, \citenamefont {Pastore}, \citenamefont {Piarulli}, \citenamefont {Girlanda}, \citenamefont {Kievsky}, \citenamefont {Lovato}, \citenamefont {Marcucci}, \citenamefont {Pieper}, \citenamefont {Viviani},\ and\ \citenamefont {Wiringa}}]{Schiavilla:2018udt}%
  \BibitemOpen
  \bibfield  {author} {\bibinfo {author} {\bibfnamefont {R.}~\bibnamefont {Schiavilla}}, \bibinfo {author} {\bibfnamefont {A.}~\bibnamefont {Baroni}}, \bibinfo {author} {\bibfnamefont {S.}~\bibnamefont {Pastore}}, \bibinfo {author} {\bibfnamefont {M.}~\bibnamefont {Piarulli}}, \bibinfo {author} {\bibfnamefont {L.}~\bibnamefont {Girlanda}}, \bibinfo {author} {\bibfnamefont {A.}~\bibnamefont {Kievsky}}, \bibinfo {author} {\bibfnamefont {A.}~\bibnamefont {Lovato}}, \bibinfo {author} {\bibfnamefont {L.~E.}\ \bibnamefont {Marcucci}}, \bibinfo {author} {\bibfnamefont {S.~C.}\ \bibnamefont {Pieper}}, \bibinfo {author} {\bibfnamefont {M.}~\bibnamefont {Viviani}}, \ and\ \bibinfo {author} {\bibfnamefont {R.~B.}\ \bibnamefont {Wiringa}},\ }\href {\doibase 10.1103/PhysRevC.99.034005} {\bibfield  {journal} {\bibinfo  {journal} {Phys. Rev. C}\ }\textbf {\bibinfo {volume} {99}},\ \bibinfo {pages} {034005} (\bibinfo {year} {2019})}\BibitemShut {NoStop}%
\bibitem [{\citenamefont {Hiyama}\ and\ \citenamefont {Suzuki}(2024)}]{Hiyama:2024zfd}%
  \BibitemOpen
  \bibfield  {author} {\bibinfo {author} {\bibfnamefont {E.}~\bibnamefont {Hiyama}}\ and\ \bibinfo {author} {\bibfnamefont {T.}~\bibnamefont {Suzuki}},\ }\href {\doibase 10.1093/ptep/ptae126} {\bibfield  {journal} {\bibinfo  {journal} {PTEP}\ }\textbf {\bibinfo {volume} {2024}},\ \bibinfo {pages} {083D02} (\bibinfo {year} {2024})},\ \Eprint {http://arxiv.org/abs/2406.17394} {arXiv:2406.17394 [nucl-th]} \BibitemShut {NoStop}%
\bibitem [{\citenamefont {Caprio}\ \emph {et~al.}(2024)\citenamefont {Caprio}, \citenamefont {Maris},\ and\ \citenamefont {Fasano}}]{Caprio:2024tzt}%
  \BibitemOpen
  \bibfield  {author} {\bibinfo {author} {\bibfnamefont {M.~A.}\ \bibnamefont {Caprio}}, \bibinfo {author} {\bibfnamefont {P.}~\bibnamefont {Maris}}, \ and\ \bibinfo {author} {\bibfnamefont {P.~J.}\ \bibnamefont {Fasano}},\ }\href@noop {} {\  (\bibinfo {year} {2024})},\ \Eprint {http://arxiv.org/abs/2409.03926} {arXiv:2409.03926 [nucl-th]} \BibitemShut {NoStop}%
\bibitem [{\citenamefont {Caprio}\ \emph {et~al.}(2025)\citenamefont {Caprio}, \citenamefont {Fasano},\ and\ \citenamefont {Maris}}]{Caprio:2025osg}%
  \BibitemOpen
  \bibfield  {author} {\bibinfo {author} {\bibfnamefont {M.~A.}\ \bibnamefont {Caprio}}, \bibinfo {author} {\bibfnamefont {P.~J.}\ \bibnamefont {Fasano}}, \ and\ \bibinfo {author} {\bibfnamefont {P.}~\bibnamefont {Maris}},\ }\href@noop {} {\  (\bibinfo {year} {2025})},\ \Eprint {http://arxiv.org/abs/2501.09228} {arXiv:2501.09228 [nucl-th]} \BibitemShut {NoStop}%
\bibitem [{\citenamefont {Shen}\ \emph {et~al.}(2024)\citenamefont {Shen}, \citenamefont {Elhatisari}, \citenamefont {Lee}, \citenamefont {Mei\ss{}ner},\ and\ \citenamefont {Ren}}]{Shen:2024qzi}%
  \BibitemOpen
  \bibfield  {author} {\bibinfo {author} {\bibfnamefont {S.}~\bibnamefont {Shen}}, \bibinfo {author} {\bibfnamefont {S.}~\bibnamefont {Elhatisari}}, \bibinfo {author} {\bibfnamefont {D.}~\bibnamefont {Lee}}, \bibinfo {author} {\bibfnamefont {U.-G.}\ \bibnamefont {Mei\ss{}ner}}, \ and\ \bibinfo {author} {\bibfnamefont {Z.}~\bibnamefont {Ren}},\ }\href@noop {} {\  (\bibinfo {year} {2024})},\ \Eprint {http://arxiv.org/abs/2411.14935} {arXiv:2411.14935 [nucl-th]} \BibitemShut {NoStop}%
\bibitem [{\citenamefont {Wolfgruber}\ \emph {et~al.}(2025)\citenamefont {Wolfgruber}, \citenamefont {Gesser}, \citenamefont {Kn\"oll}, \citenamefont {Maris},\ and\ \citenamefont {Roth}}]{Wolfgruber:2025rys}%
  \BibitemOpen
  \bibfield  {author} {\bibinfo {author} {\bibfnamefont {T.}~\bibnamefont {Wolfgruber}}, \bibinfo {author} {\bibfnamefont {T.}~\bibnamefont {Gesser}}, \bibinfo {author} {\bibfnamefont {M.}~\bibnamefont {Kn\"oll}}, \bibinfo {author} {\bibfnamefont {P.}~\bibnamefont {Maris}}, \ and\ \bibinfo {author} {\bibfnamefont {R.}~\bibnamefont {Roth}},\ }\href@noop {} {\  (\bibinfo {year} {2025})},\ \Eprint {http://arxiv.org/abs/2503.20764} {arXiv:2503.20764 [nucl-th]} \BibitemShut {NoStop}%
\bibitem [{\citenamefont {Nevo~Dinur}\ \emph {et~al.}(2019)\citenamefont {Nevo~Dinur}, \citenamefont {Hernandez}, \citenamefont {Bacca}, \citenamefont {Barnea}, \citenamefont {Ji}, \citenamefont {Pastore}, \citenamefont {Piarulli},\ and\ \citenamefont {Wiringa}}]{Nevo_Dinur_2019}%
  \BibitemOpen
  \bibfield  {author} {\bibinfo {author} {\bibfnamefont {N.}~\bibnamefont {Nevo~Dinur}}, \bibinfo {author} {\bibfnamefont {O.~J.}\ \bibnamefont {Hernandez}}, \bibinfo {author} {\bibfnamefont {S.}~\bibnamefont {Bacca}}, \bibinfo {author} {\bibfnamefont {N.}~\bibnamefont {Barnea}}, \bibinfo {author} {\bibfnamefont {C.}~\bibnamefont {Ji}}, \bibinfo {author} {\bibfnamefont {S.}~\bibnamefont {Pastore}}, \bibinfo {author} {\bibfnamefont {M.}~\bibnamefont {Piarulli}}, \ and\ \bibinfo {author} {\bibfnamefont {R.~B.}\ \bibnamefont {Wiringa}},\ }\href {\doibase 10.1103/physrevc.99.034004} {\bibfield  {journal} {\bibinfo  {journal} {Physical Review C}\ }\textbf {\bibinfo {volume} {99}} (\bibinfo {year} {2019}),\ 10.1103/physrevc.99.034004}\BibitemShut {NoStop}%
\bibitem [{\citenamefont {Dickopf}\ \emph {et~al.}(2024)\citenamefont {Dickopf}, \citenamefont {Sikora}, \citenamefont {Kaiser}, \citenamefont {Müller}, \citenamefont {Ulmer}, \citenamefont {Yerokhin}, \citenamefont {Harman}, \citenamefont {Keitel}, \citenamefont {Mooser},\ and\ \citenamefont {Blaum}}]{Dickopf_2024}%
  \BibitemOpen
  \bibfield  {author} {\bibinfo {author} {\bibfnamefont {S.}~\bibnamefont {Dickopf}}, \bibinfo {author} {\bibfnamefont {B.}~\bibnamefont {Sikora}}, \bibinfo {author} {\bibfnamefont {A.}~\bibnamefont {Kaiser}}, \bibinfo {author} {\bibfnamefont {M.}~\bibnamefont {Müller}}, \bibinfo {author} {\bibfnamefont {S.}~\bibnamefont {Ulmer}}, \bibinfo {author} {\bibfnamefont {V.~A.}\ \bibnamefont {Yerokhin}}, \bibinfo {author} {\bibfnamefont {Z.}~\bibnamefont {Harman}}, \bibinfo {author} {\bibfnamefont {C.~H.}\ \bibnamefont {Keitel}}, \bibinfo {author} {\bibfnamefont {A.}~\bibnamefont {Mooser}}, \ and\ \bibinfo {author} {\bibfnamefont {K.}~\bibnamefont {Blaum}},\ }\href {\doibase 10.1038/s41586-024-07795-1} {\bibfield  {journal} {\bibinfo  {journal} {Nature}\ }\textbf {\bibinfo {volume} {632}},\ \bibinfo {pages} {757–761} (\bibinfo {year} {2024})}\BibitemShut {NoStop}%
\bibitem [{\citenamefont {Wiringa}(1991)}]{Wiringa:1991kp}%
  \BibitemOpen
  \bibfield  {author} {\bibinfo {author} {\bibfnamefont {R.~B.}\ \bibnamefont {Wiringa}},\ }\href {\doibase 10.1103/PhysRevC.43.1585} {\bibfield  {journal} {\bibinfo  {journal} {Phys. Rev.}\ }\textbf {\bibinfo {volume} {C43}},\ \bibinfo {pages} {1585} (\bibinfo {year} {1991})}\BibitemShut {NoStop}%
\bibitem [{\citenamefont {Pudliner}\ \emph {et~al.}(1997)\citenamefont {Pudliner}, \citenamefont {Pandharipande}, \citenamefont {Carlson}, \citenamefont {Pieper},\ and\ \citenamefont {Wiringa}}]{Pudliner:1997ck}%
  \BibitemOpen
  \bibfield  {author} {\bibinfo {author} {\bibfnamefont {B.~S.}\ \bibnamefont {Pudliner}}, \bibinfo {author} {\bibfnamefont {V.~R.}\ \bibnamefont {Pandharipande}}, \bibinfo {author} {\bibfnamefont {J.}~\bibnamefont {Carlson}}, \bibinfo {author} {\bibfnamefont {S.~C.}\ \bibnamefont {Pieper}}, \ and\ \bibinfo {author} {\bibfnamefont {R.~B.}\ \bibnamefont {Wiringa}},\ }\href {\doibase 10.1103/PhysRevC.56.1720} {\bibfield  {journal} {\bibinfo  {journal} {Phys. Rev.}\ }\textbf {\bibinfo {volume} {C56}},\ \bibinfo {pages} {1720} (\bibinfo {year} {1997})},\ \Eprint {http://arxiv.org/abs/nucl-th/9705009} {arXiv:nucl-th/9705009 [nucl-th]} \BibitemShut {NoStop}%
\bibitem [{\citenamefont {Carlson}\ \emph {et~al.}(2015)\citenamefont {Carlson}, \citenamefont {Gandolfi}, \citenamefont {Pederiva}, \citenamefont {Pieper}, \citenamefont {Schiavilla}, \citenamefont {Schmidt},\ and\ \citenamefont {Wiringa}}]{Carlson:2014vla}%
  \BibitemOpen
  \bibfield  {author} {\bibinfo {author} {\bibfnamefont {J.}~\bibnamefont {Carlson}}, \bibinfo {author} {\bibfnamefont {S.}~\bibnamefont {Gandolfi}}, \bibinfo {author} {\bibfnamefont {F.}~\bibnamefont {Pederiva}}, \bibinfo {author} {\bibfnamefont {S.~C.}\ \bibnamefont {Pieper}}, \bibinfo {author} {\bibfnamefont {R.}~\bibnamefont {Schiavilla}}, \bibinfo {author} {\bibfnamefont {K.~E.}\ \bibnamefont {Schmidt}}, \ and\ \bibinfo {author} {\bibfnamefont {R.~B.}\ \bibnamefont {Wiringa}},\ }\href {\doibase 10.1103/RevModPhys.87.1067} {\bibfield  {journal} {\bibinfo  {journal} {Rev. Mod. Phys.}\ }\textbf {\bibinfo {volume} {87}},\ \bibinfo {pages} {1067} (\bibinfo {year} {2015})},\ \Eprint {http://arxiv.org/abs/1412.3081} {arXiv:1412.3081 [nucl-th]} \BibitemShut {NoStop}%
\bibitem [{\citenamefont {Gandolfi}\ \emph {et~al.}(2020)\citenamefont {Gandolfi}, \citenamefont {Lonardoni}, \citenamefont {Lovato},\ and\ \citenamefont {Piarulli}}]{Gandolfi:2020pbj}%
  \BibitemOpen
  \bibfield  {author} {\bibinfo {author} {\bibfnamefont {S.}~\bibnamefont {Gandolfi}}, \bibinfo {author} {\bibfnamefont {D.}~\bibnamefont {Lonardoni}}, \bibinfo {author} {\bibfnamefont {A.}~\bibnamefont {Lovato}}, \ and\ \bibinfo {author} {\bibfnamefont {M.}~\bibnamefont {Piarulli}},\ }\href@noop {} {\  (\bibinfo {year} {2020})},\ \Eprint {http://arxiv.org/abs/2001.01374} {arXiv:2001.01374 [nucl-th]} \BibitemShut {NoStop}%
\bibitem [{\citenamefont {Park}\ \emph {et~al.}(1996)\citenamefont {Park}, \citenamefont {Min},\ and\ \citenamefont {Rho}}]{Park:1995pn}%
  \BibitemOpen
  \bibfield  {author} {\bibinfo {author} {\bibfnamefont {T.-S.}\ \bibnamefont {Park}}, \bibinfo {author} {\bibfnamefont {D.-P.}\ \bibnamefont {Min}}, \ and\ \bibinfo {author} {\bibfnamefont {M.}~\bibnamefont {Rho}},\ }\href {\doibase 10.1016/0375-9474(95)00406-8} {\bibfield  {journal} {\bibinfo  {journal} {Nucl. Phys.}\ }\textbf {\bibinfo {volume} {A596}},\ \bibinfo {pages} {515} (\bibinfo {year} {1996})},\ \Eprint {http://arxiv.org/abs/nucl-th/9505017} {arXiv:nucl-th/9505017 [nucl-th]} \BibitemShut {NoStop}%
\bibitem [{\citenamefont {Walzl}\ and\ \citenamefont {Meißner}(2001)}]{WALZL200137}%
  \BibitemOpen
  \bibfield  {author} {\bibinfo {author} {\bibfnamefont {M.}~\bibnamefont {Walzl}}\ and\ \bibinfo {author} {\bibfnamefont {U.-G.}\ \bibnamefont {Meißner}},\ }\href {\doibase https://doi.org/10.1016/S0370-2693(01)00727-4} {\bibfield  {journal} {\bibinfo  {journal} {Physics Letters B}\ }\textbf {\bibinfo {volume} {513}},\ \bibinfo {pages} {37} (\bibinfo {year} {2001})}\BibitemShut {NoStop}%
\bibitem [{\citenamefont {Phillips}(2003)}]{PHILLIPS200312}%
  \BibitemOpen
  \bibfield  {author} {\bibinfo {author} {\bibfnamefont {D.~R.}\ \bibnamefont {Phillips}},\ }\href {\doibase https://doi.org/10.1016/S0370-2693(03)00867-0} {\bibfield  {journal} {\bibinfo  {journal} {Physics Letters B}\ }\textbf {\bibinfo {volume} {567}},\ \bibinfo {pages} {12} (\bibinfo {year} {2003})}\BibitemShut {NoStop}%
\bibitem [{\citenamefont {Phillips}(2007)}]{Phillips_2007}%
  \BibitemOpen
  \bibfield  {author} {\bibinfo {author} {\bibfnamefont {D.~R.}\ \bibnamefont {Phillips}},\ }\href {\doibase 10.1088/0954-3899/34/2/015} {\bibfield  {journal} {\bibinfo  {journal} {Journal of Physics G: Nuclear and Particle Physics}\ }\textbf {\bibinfo {volume} {34}},\ \bibinfo {pages} {365} (\bibinfo {year} {2007})}\BibitemShut {NoStop}%
\bibitem [{\citenamefont {Kolling}\ \emph {et~al.}(2009)\citenamefont {Kolling}, \citenamefont {Epelbaum}, \citenamefont {Krebs},\ and\ \citenamefont {Meissner}}]{Kolling:2009iq}%
  \BibitemOpen
  \bibfield  {author} {\bibinfo {author} {\bibfnamefont {S.}~\bibnamefont {Kolling}}, \bibinfo {author} {\bibfnamefont {E.}~\bibnamefont {Epelbaum}}, \bibinfo {author} {\bibfnamefont {H.}~\bibnamefont {Krebs}}, \ and\ \bibinfo {author} {\bibfnamefont {U.~G.}\ \bibnamefont {Meissner}},\ }\href {\doibase 10.1103/PhysRevC.80.045502} {\bibfield  {journal} {\bibinfo  {journal} {Phys. Rev.}\ }\textbf {\bibinfo {volume} {C80}},\ \bibinfo {pages} {045502} (\bibinfo {year} {2009})},\ \Eprint {http://arxiv.org/abs/0907.3437} {arXiv:0907.3437 [nucl-th]} \BibitemShut {NoStop}%
\bibitem [{\citenamefont {Kolling}\ \emph {et~al.}(2011)\citenamefont {Kolling}, \citenamefont {Epelbaum}, \citenamefont {Krebs},\ and\ \citenamefont {Meissner}}]{Kolling:2011mt}%
  \BibitemOpen
  \bibfield  {author} {\bibinfo {author} {\bibfnamefont {S.}~\bibnamefont {Kolling}}, \bibinfo {author} {\bibfnamefont {E.}~\bibnamefont {Epelbaum}}, \bibinfo {author} {\bibfnamefont {H.}~\bibnamefont {Krebs}}, \ and\ \bibinfo {author} {\bibfnamefont {U.~G.}\ \bibnamefont {Meissner}},\ }\href {\doibase 10.1103/PhysRevC.84.054008} {\bibfield  {journal} {\bibinfo  {journal} {Phys. Rev.}\ }\textbf {\bibinfo {volume} {C84}},\ \bibinfo {pages} {054008} (\bibinfo {year} {2011})},\ \Eprint {http://arxiv.org/abs/1107.0602} {arXiv:1107.0602 [nucl-th]} \BibitemShut {NoStop}%
\bibitem [{\citenamefont {Krebs}\ \emph {et~al.}(2017)\citenamefont {Krebs}, \citenamefont {Epelbaum},\ and\ \citenamefont {Meißner}}]{Krebs:2016rqz}%
  \BibitemOpen
  \bibfield  {author} {\bibinfo {author} {\bibfnamefont {H.}~\bibnamefont {Krebs}}, \bibinfo {author} {\bibfnamefont {E.}~\bibnamefont {Epelbaum}}, \ and\ \bibinfo {author} {\bibfnamefont {U.~G.}\ \bibnamefont {Meißner}},\ }\href {\doibase 10.1016/j.aop.2017.01.021} {\bibfield  {journal} {\bibinfo  {journal} {Annals Phys.}\ }\textbf {\bibinfo {volume} {378}},\ \bibinfo {pages} {317} (\bibinfo {year} {2017})},\ \Eprint {http://arxiv.org/abs/1610.03569} {arXiv:1610.03569 [nucl-th]} \BibitemShut {NoStop}%
\bibitem [{\citenamefont {Krebs}\ \emph {et~al.}(2019)\citenamefont {Krebs}, \citenamefont {Epelbaum},\ and\ \citenamefont {Mei\ss{}ner}}]{Krebs:2019aka}%
  \BibitemOpen
  \bibfield  {author} {\bibinfo {author} {\bibfnamefont {H.}~\bibnamefont {Krebs}}, \bibinfo {author} {\bibfnamefont {E.}~\bibnamefont {Epelbaum}}, \ and\ \bibinfo {author} {\bibfnamefont {U.~G.}\ \bibnamefont {Mei\ss{}ner}},\ }\href {\doibase 10.1007/s00601-019-1500-5} {\bibfield  {journal} {\bibinfo  {journal} {Few Body Syst.}\ }\textbf {\bibinfo {volume} {60}},\ \bibinfo {pages} {31} (\bibinfo {year} {2019})},\ \Eprint {http://arxiv.org/abs/1902.06839} {arXiv:1902.06839 [nucl-th]} \BibitemShut {NoStop}%
\bibitem [{\citenamefont {Gnech}\ and\ \citenamefont {Schiavilla}(2022)}]{Gnech:2022vwr}%
  \BibitemOpen
  \bibfield  {author} {\bibinfo {author} {\bibfnamefont {A.}~\bibnamefont {Gnech}}\ and\ \bibinfo {author} {\bibfnamefont {R.}~\bibnamefont {Schiavilla}},\ }\href {\doibase 10.1103/PhysRevC.106.044001} {\bibfield  {journal} {\bibinfo  {journal} {Phys. Rev. C}\ }\textbf {\bibinfo {volume} {106}},\ \bibinfo {pages} {044001} (\bibinfo {year} {2022})},\ \Eprint {http://arxiv.org/abs/2207.05528} {arXiv:2207.05528 [nucl-th]} \BibitemShut {NoStop}%
\bibitem [{\citenamefont {Carlson}\ and\ \citenamefont {Schiavilla}(1998)}]{Carlson:1997qn}%
  \BibitemOpen
  \bibfield  {author} {\bibinfo {author} {\bibfnamefont {J.}~\bibnamefont {Carlson}}\ and\ \bibinfo {author} {\bibfnamefont {R.}~\bibnamefont {Schiavilla}},\ }\href {\doibase 10.1103/RevModPhys.70.743} {\bibfield  {journal} {\bibinfo  {journal} {Rev. Mod. Phys.}\ }\textbf {\bibinfo {volume} {70}},\ \bibinfo {pages} {743} (\bibinfo {year} {1998})}\BibitemShut {NoStop}%
\bibitem [{\citenamefont {Kelly}(2004)}]{Kelly:2004}%
  \BibitemOpen
  \bibfield  {author} {\bibinfo {author} {\bibfnamefont {J.~J.}\ \bibnamefont {Kelly}},\ }\href {\doibase 10.1103/PhysRevC.70.068202} {\bibfield  {journal} {\bibinfo  {journal} {Phys. Rev. C}\ }\textbf {\bibinfo {volume} {70}},\ \bibinfo {pages} {068202} (\bibinfo {year} {2004})}\BibitemShut {NoStop}%
\bibitem [{\citenamefont {Suda}(2022)}]{Suda:2022hsm}%
  \BibitemOpen
  \bibfield  {author} {\bibinfo {author} {\bibfnamefont {T.}~\bibnamefont {Suda}},\ }\href {\doibase 10.1088/1742-6596/2391/1/012004} {\bibfield  {journal} {\bibinfo  {journal} {J. Phys. Conf. Ser.}\ }\textbf {\bibinfo {volume} {2391}},\ \bibinfo {pages} {012004} (\bibinfo {year} {2022})}\BibitemShut {NoStop}%
\bibitem [{\citenamefont {Gasparian}\ \emph {et~al.}(2020)\citenamefont {Gasparian} \emph {et~al.}}]{PRad:2020oor}%
  \BibitemOpen
  \bibfield  {author} {\bibinfo {author} {\bibfnamefont {A.}~\bibnamefont {Gasparian}} \emph {et~al.} (\bibinfo {collaboration} {PRad}),\ }\href@noop {} {\  (\bibinfo {year} {2020})},\ \Eprint {http://arxiv.org/abs/2009.10510} {arXiv:2009.10510 [nucl-ex]} \BibitemShut {NoStop}%
\bibitem [{\citenamefont {Wang}\ \emph {et~al.}(2022)\citenamefont {Wang} \emph {et~al.}}]{A1:2022wzx}%
  \BibitemOpen
  \bibfield  {author} {\bibinfo {author} {\bibfnamefont {Y.}~\bibnamefont {Wang}} \emph {et~al.} (\bibinfo {collaboration} {A1, MAGIX}),\ }\href {\doibase 10.1103/PhysRevC.106.044610} {\bibfield  {journal} {\bibinfo  {journal} {Phys. Rev. C}\ }\textbf {\bibinfo {volume} {106}},\ \bibinfo {pages} {044610} (\bibinfo {year} {2022})},\ \Eprint {http://arxiv.org/abs/2208.13689} {arXiv:2208.13689 [nucl-ex]} \BibitemShut {NoStop}%
\bibitem [{\citenamefont {Cline}\ \emph {et~al.}(2021)\citenamefont {Cline}, \citenamefont {Bernauer}, \citenamefont {Downie},\ and\ \citenamefont {Gilman}}]{Cline:2021ehf}%
  \BibitemOpen
  \bibfield  {author} {\bibinfo {author} {\bibfnamefont {E.}~\bibnamefont {Cline}}, \bibinfo {author} {\bibfnamefont {J.}~\bibnamefont {Bernauer}}, \bibinfo {author} {\bibfnamefont {E.~J.}\ \bibnamefont {Downie}}, \ and\ \bibinfo {author} {\bibfnamefont {R.}~\bibnamefont {Gilman}},\ }\href {\doibase 10.21468/SciPostPhysProc.5.023} {\bibfield  {journal} {\bibinfo  {journal} {SciPost Phys. Proc.}\ }\textbf {\bibinfo {volume} {5}},\ \bibinfo {pages} {023} (\bibinfo {year} {2021})}\BibitemShut {NoStop}%
\bibitem [{\citenamefont {Friedrich}(2024)}]{Friedrich:2024ylw}%
  \BibitemOpen
  \bibfield  {author} {\bibinfo {author} {\bibfnamefont {J.~M.}\ \bibnamefont {Friedrich}} (\bibinfo {collaboration} {AMBER}),\ }\href {\doibase 10.1051/epjconf/202430306001} {\bibfield  {journal} {\bibinfo  {journal} {EPJ Web Conf.}\ }\textbf {\bibinfo {volume} {303}},\ \bibinfo {pages} {06001} (\bibinfo {year} {2024})}\BibitemShut {NoStop}%
\bibitem [{\citenamefont {Filin}\ \emph {et~al.}(2020)\citenamefont {Filin}, \citenamefont {Baru}, \citenamefont {Epelbaum}, \citenamefont {Krebs}, \citenamefont {M\"oller},\ and\ \citenamefont {Reinert}}]{Filin:2019eoe}%
  \BibitemOpen
  \bibfield  {author} {\bibinfo {author} {\bibfnamefont {A.~A.}\ \bibnamefont {Filin}}, \bibinfo {author} {\bibfnamefont {V.}~\bibnamefont {Baru}}, \bibinfo {author} {\bibfnamefont {E.}~\bibnamefont {Epelbaum}}, \bibinfo {author} {\bibfnamefont {H.}~\bibnamefont {Krebs}}, \bibinfo {author} {\bibfnamefont {D.}~\bibnamefont {M\"oller}}, \ and\ \bibinfo {author} {\bibfnamefont {P.}~\bibnamefont {Reinert}},\ }\href {\doibase 10.1103/PhysRevLett.124.082501} {\bibfield  {journal} {\bibinfo  {journal} {Phys. Rev. Lett.}\ }\textbf {\bibinfo {volume} {124}},\ \bibinfo {pages} {082501} (\bibinfo {year} {2020})},\ \Eprint {http://arxiv.org/abs/1911.04877} {arXiv:1911.04877 [nucl-th]} \BibitemShut {NoStop}%
\bibitem [{\citenamefont {Filin}\ \emph {et~al.}(2021)\citenamefont {Filin}, \citenamefont {M\"oller}, \citenamefont {Baru}, \citenamefont {Epelbaum}, \citenamefont {Krebs},\ and\ \citenamefont {Reinert}}]{Filin:2020tcs}%
  \BibitemOpen
  \bibfield  {author} {\bibinfo {author} {\bibfnamefont {A.~A.}\ \bibnamefont {Filin}}, \bibinfo {author} {\bibfnamefont {D.}~\bibnamefont {M\"oller}}, \bibinfo {author} {\bibfnamefont {V.}~\bibnamefont {Baru}}, \bibinfo {author} {\bibfnamefont {E.}~\bibnamefont {Epelbaum}}, \bibinfo {author} {\bibfnamefont {H.}~\bibnamefont {Krebs}}, \ and\ \bibinfo {author} {\bibfnamefont {P.}~\bibnamefont {Reinert}},\ }\href {\doibase 10.1103/PhysRevC.103.024313} {\bibfield  {journal} {\bibinfo  {journal} {Phys. Rev. C}\ }\textbf {\bibinfo {volume} {103}},\ \bibinfo {pages} {024313} (\bibinfo {year} {2021})},\ \Eprint {http://arxiv.org/abs/2009.08911} {arXiv:2009.08911 [nucl-th]} \BibitemShut {NoStop}%
\bibitem [{\citenamefont {Pastore}\ \emph {et~al.}(2013)\citenamefont {Pastore}, \citenamefont {Pieper}, \citenamefont {Schiavilla},\ and\ \citenamefont {Wiringa}}]{Pastore:2012rp}%
  \BibitemOpen
  \bibfield  {author} {\bibinfo {author} {\bibfnamefont {S.}~\bibnamefont {Pastore}}, \bibinfo {author} {\bibfnamefont {S.~C.}\ \bibnamefont {Pieper}}, \bibinfo {author} {\bibfnamefont {R.}~\bibnamefont {Schiavilla}}, \ and\ \bibinfo {author} {\bibfnamefont {R.~B.}\ \bibnamefont {Wiringa}},\ }\href {\doibase 10.1103/PhysRevC.87.035503} {\bibfield  {journal} {\bibinfo  {journal} {Phys. Rev.}\ }\textbf {\bibinfo {volume} {C87}},\ \bibinfo {pages} {035503} (\bibinfo {year} {2013})},\ \Eprint {http://arxiv.org/abs/1212.3375} {arXiv:1212.3375 [nucl-th]} \BibitemShut {NoStop}%
\bibitem [{\citenamefont {King}\ and\ \citenamefont {Pastore}(2024)}]{King:2024zbv}%
  \BibitemOpen
  \bibfield  {author} {\bibinfo {author} {\bibfnamefont {G.~B.}\ \bibnamefont {King}}\ and\ \bibinfo {author} {\bibfnamefont {S.}~\bibnamefont {Pastore}},\ }\href@noop {} {\  (\bibinfo {year} {2024})},\ \Eprint {http://arxiv.org/abs/2402.06602} {arXiv:2402.06602 [nucl-th]} \BibitemShut {NoStop}%
\bibitem [{\citenamefont {Amroun}\ \emph {et~al.}(1994)\citenamefont {Amroun} \emph {et~al.}}]{Amroun:1994qj}%
  \BibitemOpen
  \bibfield  {author} {\bibinfo {author} {\bibfnamefont {A.}~\bibnamefont {Amroun}} \emph {et~al.},\ }\href {\doibase 10.1016/0375-9474(94)90925-3} {\bibfield  {journal} {\bibinfo  {journal} {Nucl. Phys.}\ }\textbf {\bibinfo {volume} {A579}},\ \bibinfo {pages} {596} (\bibinfo {year} {1994})}\BibitemShut {NoStop}%
\bibitem [{\citenamefont {Shiner}\ \emph {et~al.}(1995)\citenamefont {Shiner}, \citenamefont {Dixson},\ and\ \citenamefont {Vedantham}}]{Shiner:1995zz}%
  \BibitemOpen
  \bibfield  {author} {\bibinfo {author} {\bibfnamefont {D.}~\bibnamefont {Shiner}}, \bibinfo {author} {\bibfnamefont {R.}~\bibnamefont {Dixson}}, \ and\ \bibinfo {author} {\bibfnamefont {V.}~\bibnamefont {Vedantham}},\ }\href {\doibase 10.1103/PhysRevLett.74.3553} {\bibfield  {journal} {\bibinfo  {journal} {Phys. Rev. Lett.}\ }\textbf {\bibinfo {volume} {74}},\ \bibinfo {pages} {3553} (\bibinfo {year} {1995})}\BibitemShut {NoStop}%
\bibitem [{\citenamefont {Krauth}\ \emph {et~al.}(2021)\citenamefont {Krauth} \emph {et~al.}}]{Krauth:2021foz}%
  \BibitemOpen
  \bibfield  {author} {\bibinfo {author} {\bibfnamefont {J.~J.}\ \bibnamefont {Krauth}} \emph {et~al.},\ }\href {\doibase 10.1038/s41586-021-03183-1} {\bibfield  {journal} {\bibinfo  {journal} {Nature}\ }\textbf {\bibinfo {volume} {589}},\ \bibinfo {pages} {527} (\bibinfo {year} {2021})}\BibitemShut {NoStop}%
\bibitem [{\citenamefont {Lu}\ \emph {et~al.}(2013)\citenamefont {Lu}, \citenamefont {Mueller}, \citenamefont {Drake}, \citenamefont {Noertershaeuser}, \citenamefont {Pieper},\ and\ \citenamefont {Yan}}]{Lu:2013ena}%
  \BibitemOpen
  \bibfield  {author} {\bibinfo {author} {\bibfnamefont {Z.~T.}\ \bibnamefont {Lu}}, \bibinfo {author} {\bibfnamefont {P.}~\bibnamefont {Mueller}}, \bibinfo {author} {\bibfnamefont {G.~W.~F.}\ \bibnamefont {Drake}}, \bibinfo {author} {\bibfnamefont {W.}~\bibnamefont {Noertershaeuser}}, \bibinfo {author} {\bibfnamefont {S.~C.}\ \bibnamefont {Pieper}}, \ and\ \bibinfo {author} {\bibfnamefont {Z.~C.}\ \bibnamefont {Yan}},\ }\href {\doibase 10.1103/RevModPhys.85.1383} {\bibfield  {journal} {\bibinfo  {journal} {Rev. Mod. Phys.}\ }\textbf {\bibinfo {volume} {85}},\ \bibinfo {pages} {1383} (\bibinfo {year} {2013})},\ \Eprint {http://arxiv.org/abs/1307.2872} {arXiv:1307.2872 [nucl-ex]} \BibitemShut {NoStop}%
\bibitem [{\citenamefont {Nortershauser}\ \emph {et~al.}(2011)\citenamefont {Nortershauser}, \citenamefont {Neff}, \citenamefont {Sanchez},\ and\ \citenamefont {Sick}}]{Nortershauser:2011zz}%
  \BibitemOpen
  \bibfield  {author} {\bibinfo {author} {\bibfnamefont {W.}~\bibnamefont {Nortershauser}}, \bibinfo {author} {\bibfnamefont {T.}~\bibnamefont {Neff}}, \bibinfo {author} {\bibfnamefont {R.}~\bibnamefont {Sanchez}}, \ and\ \bibinfo {author} {\bibfnamefont {I.}~\bibnamefont {Sick}},\ }\href {\doibase 10.1103/PhysRevC.84.024307} {\bibfield  {journal} {\bibinfo  {journal} {Phys. Rev.}\ }\textbf {\bibinfo {volume} {C84}},\ \bibinfo {pages} {024307} (\bibinfo {year} {2011})}\BibitemShut {NoStop}%
\bibitem [{\citenamefont {Nortershauser}\ \emph {et~al.}(2009)\citenamefont {Nortershauser} \emph {et~al.}}]{Nortershauser:2008vp}%
  \BibitemOpen
  \bibfield  {author} {\bibinfo {author} {\bibfnamefont {W.}~\bibnamefont {Nortershauser}} \emph {et~al.},\ }\href {\doibase 10.1103/PhysRevLett.102.062503} {\bibfield  {journal} {\bibinfo  {journal} {Phys. Rev. Lett.}\ }\textbf {\bibinfo {volume} {102}},\ \bibinfo {pages} {062503} (\bibinfo {year} {2009})},\ \Eprint {http://arxiv.org/abs/0809.2607} {arXiv:0809.2607 [nucl-ex]} \BibitemShut {NoStop}%
\bibitem [{\citenamefont {Sick}(2008)}]{Sick:2008zza}%
  \BibitemOpen
  \bibfield  {author} {\bibinfo {author} {\bibfnamefont {I.}~\bibnamefont {Sick}},\ }\href {\doibase 10.1103/PhysRevC.77.041302} {\bibfield  {journal} {\bibinfo  {journal} {Phys. Rev. C}\ }\textbf {\bibinfo {volume} {77}},\ \bibinfo {pages} {041302} (\bibinfo {year} {2008})}\BibitemShut {NoStop}%
\bibitem [{\citenamefont {Jansen}\ \emph {et~al.}(1972)\citenamefont {Jansen}, \citenamefont {Peerdeman},\ and\ \citenamefont {De~Vries}}]{Jansen:1972iui}%
  \BibitemOpen
  \bibfield  {author} {\bibinfo {author} {\bibfnamefont {J.~A.}\ \bibnamefont {Jansen}}, \bibinfo {author} {\bibfnamefont {R.~T.}\ \bibnamefont {Peerdeman}}, \ and\ \bibinfo {author} {\bibfnamefont {C.}~\bibnamefont {De~Vries}},\ }\href {\doibase 10.1016/0375-9474(72)90062-0} {\bibfield  {journal} {\bibinfo  {journal} {Nucl. Phys. A}\ }\textbf {\bibinfo {volume} {188}},\ \bibinfo {pages} {337} (\bibinfo {year} {1972})}\BibitemShut {NoStop}%
\bibitem [{\citenamefont {Cichocki}\ \emph {et~al.}(1995)\citenamefont {Cichocki}, \citenamefont {Dubach}, \citenamefont {Hicks}, \citenamefont {Peterson}, \citenamefont {de~Jager}, \citenamefont {de~Vries}, \citenamefont {Kalantar-Nayestanaki},\ and\ \citenamefont {Sato}}]{Cichocki:1995zz}%
  \BibitemOpen
  \bibfield  {author} {\bibinfo {author} {\bibfnamefont {A.}~\bibnamefont {Cichocki}}, \bibinfo {author} {\bibfnamefont {J.}~\bibnamefont {Dubach}}, \bibinfo {author} {\bibfnamefont {R.~S.}\ \bibnamefont {Hicks}}, \bibinfo {author} {\bibfnamefont {G.~A.}\ \bibnamefont {Peterson}}, \bibinfo {author} {\bibfnamefont {C.~W.}\ \bibnamefont {de~Jager}}, \bibinfo {author} {\bibfnamefont {H.}~\bibnamefont {de~Vries}}, \bibinfo {author} {\bibfnamefont {N.}~\bibnamefont {Kalantar-Nayestanaki}}, \ and\ \bibinfo {author} {\bibfnamefont {T.}~\bibnamefont {Sato}},\ }\href {\doibase 10.1103/PhysRevC.51.2406} {\bibfield  {journal} {\bibinfo  {journal} {Phys. Rev. C}\ }\textbf {\bibinfo {volume} {51}},\ \bibinfo {pages} {2406} (\bibinfo {year} {1995})}\BibitemShut {NoStop}%
\bibitem [{\citenamefont {Bernauer}\ \emph {et~al.}(2022)\citenamefont {Bernauer}, \citenamefont {Corliss}, \citenamefont {Gardner}, \citenamefont {Hasinoff}, \citenamefont {Kanungo}, \citenamefont {Martin}, \citenamefont {Milner}, \citenamefont {Pachal}, \citenamefont {Suda},\ and\ \citenamefont {Yen}}]{Bernauer:2022cqe}%
  \BibitemOpen
  \bibfield  {author} {\bibinfo {author} {\bibfnamefont {J.}~\bibnamefont {Bernauer}}, \bibinfo {author} {\bibfnamefont {R.}~\bibnamefont {Corliss}}, \bibinfo {author} {\bibfnamefont {S.}~\bibnamefont {Gardner}}, \bibinfo {author} {\bibfnamefont {M.}~\bibnamefont {Hasinoff}}, \bibinfo {author} {\bibfnamefont {R.}~\bibnamefont {Kanungo}}, \bibinfo {author} {\bibfnamefont {J.}~\bibnamefont {Martin}}, \bibinfo {author} {\bibfnamefont {R.}~\bibnamefont {Milner}}, \bibinfo {author} {\bibfnamefont {K.}~\bibnamefont {Pachal}}, \bibinfo {author} {\bibfnamefont {T.}~\bibnamefont {Suda}}, \ and\ \bibinfo {author} {\bibfnamefont {S.}~\bibnamefont {Yen}},\ }\href {\doibase 10.1088/1742-6596/2391/1/012001} {\bibfield  {journal} {\bibinfo  {journal} {J. Phys. Conf. Ser.}\ }\textbf {\bibinfo {volume} {2391}},\ \bibinfo {pages} {012001} (\bibinfo {year} {2022})},\ \Eprint {http://arxiv.org/abs/2305.09066} {arXiv:2305.09066 [nucl-ex]} \BibitemShut {NoStop}%
\bibitem [{\citenamefont {Tsukada}\ \emph {et~al.}(2023)\citenamefont {Tsukada}, \citenamefont {Abe}, \citenamefont {Enokizono}, \citenamefont {Goke}, \citenamefont {Hara}, \citenamefont {Honda}, \citenamefont {Hori}, \citenamefont {Ichikawa}, \citenamefont {Ito}, \citenamefont {Kurita}, \citenamefont {Legris}, \citenamefont {Maehara}, \citenamefont {Ohnishi}, \citenamefont {Ogawara}, \citenamefont {Suda}, \citenamefont {Tamae}, \citenamefont {Wakasugi}, \citenamefont {Watanabe},\ and\ \citenamefont {Wauke}}]{Tsukada:2023}%
  \BibitemOpen
  \bibfield  {author} {\bibinfo {author} {\bibfnamefont {K.}~\bibnamefont {Tsukada}}, \bibinfo {author} {\bibfnamefont {Y.}~\bibnamefont {Abe}}, \bibinfo {author} {\bibfnamefont {A.}~\bibnamefont {Enokizono}}, \bibinfo {author} {\bibfnamefont {T.}~\bibnamefont {Goke}}, \bibinfo {author} {\bibfnamefont {M.}~\bibnamefont {Hara}}, \bibinfo {author} {\bibfnamefont {Y.}~\bibnamefont {Honda}}, \bibinfo {author} {\bibfnamefont {T.}~\bibnamefont {Hori}}, \bibinfo {author} {\bibfnamefont {S.}~\bibnamefont {Ichikawa}}, \bibinfo {author} {\bibfnamefont {Y.}~\bibnamefont {Ito}}, \bibinfo {author} {\bibfnamefont {K.}~\bibnamefont {Kurita}}, \bibinfo {author} {\bibfnamefont {C.}~\bibnamefont {Legris}}, \bibinfo {author} {\bibfnamefont {Y.}~\bibnamefont {Maehara}}, \bibinfo {author} {\bibfnamefont {T.}~\bibnamefont {Ohnishi}}, \bibinfo {author} {\bibfnamefont {R.}~\bibnamefont {Ogawara}}, \bibinfo {author} {\bibfnamefont {T.}~\bibnamefont {Suda}}, \bibinfo {author} {\bibfnamefont {T.}~\bibnamefont {Tamae}}, \bibinfo
  {author} {\bibfnamefont {M.}~\bibnamefont {Wakasugi}}, \bibinfo {author} {\bibfnamefont {M.}~\bibnamefont {Watanabe}}, \ and\ \bibinfo {author} {\bibfnamefont {H.}~\bibnamefont {Wauke}},\ }\href {\doibase 10.1103/PhysRevLett.131.092502} {\bibfield  {journal} {\bibinfo  {journal} {Phys. Rev. Lett.}\ }\textbf {\bibinfo {volume} {131}},\ \bibinfo {pages} {092502} (\bibinfo {year} {2023})}\BibitemShut {NoStop}%
\bibitem [{\citenamefont {Minamisono}\ \emph {et~al.}(2013)\citenamefont {Minamisono}, \citenamefont {Mantica}, \citenamefont {Klose}, \citenamefont {Vinnikova}, \citenamefont {Schneider}, \citenamefont {Johnson},\ and\ \citenamefont {Barquest}}]{Minamisono:2013}%
  \BibitemOpen
  \bibfield  {author} {\bibinfo {author} {\bibfnamefont {K.}~\bibnamefont {Minamisono}}, \bibinfo {author} {\bibfnamefont {P.}~\bibnamefont {Mantica}}, \bibinfo {author} {\bibfnamefont {A.}~\bibnamefont {Klose}}, \bibinfo {author} {\bibfnamefont {S.}~\bibnamefont {Vinnikova}}, \bibinfo {author} {\bibfnamefont {A.}~\bibnamefont {Schneider}}, \bibinfo {author} {\bibfnamefont {B.}~\bibnamefont {Johnson}}, \ and\ \bibinfo {author} {\bibfnamefont {B.}~\bibnamefont {Barquest}},\ }\href {\doibase https://doi.org/10.1016/j.nima.2013.01.038} {\bibfield  {journal} {\bibinfo  {journal} {Nuclear Instruments and Methods in Physics Research Section A: Accelerators, Spectrometers, Detectors and Associated Equipment}\ }\textbf {\bibinfo {volume} {709}},\ \bibinfo {pages} {85} (\bibinfo {year} {2013})}\BibitemShut {NoStop}%
\bibitem [{\citenamefont {Ohayon}\ \emph {et~al.}(2024)\citenamefont {Ohayon} \emph {et~al.}}]{Ohayon:2023hze}%
  \BibitemOpen
  \bibfield  {author} {\bibinfo {author} {\bibfnamefont {B.}~\bibnamefont {Ohayon}} \emph {et~al.},\ }\href {\doibase 10.3390/physics6010015} {\bibfield  {journal} {\bibinfo  {journal} {MDPI Physics}\ }\textbf {\bibinfo {volume} {6}},\ \bibinfo {pages} {206} (\bibinfo {year} {2024})},\ \Eprint {http://arxiv.org/abs/2310.03846} {arXiv:2310.03846 [physics.atom-ph]} \BibitemShut {NoStop}%
\end{thebibliography}%

\end{document}